# A Weight of Evidence approach to classify nanomaterials according to the EU Classification, Labelling and Packaging regulation criteria

Gianpietro Basei[1*], Alex Zabeo[1], Kirsten Rasmussen[2], Georgia Tsiliki[3], Danail Hristozov[1*]

[1]GreenDecision Srl., Venezia, Italy, [2]European Commission, Joint Research Centre (JRC), Ispra, Italy, [3]ATHENA Research and Innovation Centre, Athens, Greece

* Corresponding authors <gianpietro.basei@greendecision.eu; danail.hristozov@greendecision.eu>

**Abstract:** In the context of the European Union (EU) Horizon 2020 GRACIOUS project (Grouping, Read-Across, Characterisation and classification framework for regulatory risk assessment of manufactured nanomaterials and Safer design of nano-enabled products), we proposed a quantitative Weight of Evidence (WoE) approach for hazard classification of nanomaterials (NMs). This approach is based on the requirements of the European Regulation on Classification, Labelling and Packaging of Substances and Mixtures (the CLP regulation), which implements the United Nations' Globally Harmonized System of Classification and Labelling of Chemicals (UN GHS) in the European Union. The goal of this WoE methodology is to facilitate classification of NMs according to CLP criteria, following the decision trees defined in ECHA's CLP regulatory guidance. In the WoE, results from heterogeneous studies are weighted according to data quality and completeness criteria, integrated, and then evaluated by expert judgment to obtain a hazard classification, resulting in a coherent and justifiable methodology. Moreover, the probabilistic nature of the proposed approach enables highlighting the uncertainty in the analysis. The proposed methodology involves the following stages: (1) collection of data for different NMs related to the endpoint of interest: each study related to each NM is referred as a Line of Evidence (LoE); (2) computation of weighted scores for each LoE: each LoE is weighted by a score calculated based on data quality and completeness criteria defined in the GRACIOUS project; (3) comparison and integration of the weighed LoEs for each NM: A Monte Carlo resampling approach is adopted to quantitatively and probabilistically integrate the weighted evidence; and (4) assignment of each NM to a hazard class: according to the results, each NM is assigned to one of the classes defined by the CLP regulation. Furthermore, to facilitate the integration and the classification of the weighted LoEs, an online R tool was developed. Finally, the approach was tested against an endpoint relevant to CLP (Aquatic Toxicity) using data retrieved from the eNanoMapper database, results obtained were consistent to results in REACH registration dossiers and in recent literature.

## 1 Introduction

In the last decades, engineered nanomaterials (NMs) have been increasingly used in a wide variety of industrial applications and consumer products [1], including among others: (1) manufacturing and materials, where NMs are for example integrated into clothes to make them odour-resistant, used as barriers to gases in plastic film, or as fillers of gaps between carbon fibres, (2) coatings and paints, in which NMs enhance resistance to deterioration by ultraviolet (UV) light, fungi or algae, (3) products for environmental clean-up that e.g. break down oil and other pollutants, (4) cosmetics, where NMs are added as UV-filters or colourants, (5) applications in energy and electronics, e.g. in television panels, batteries and transistors.

Indeed, nanotechnology has proven its great potential to innovate, but its widespread application has also raised societal concerns about the possible health and environmental risks of the NMs [2–4]. To address these concerns, the European funding agencies and regulators have invested in ensuring that scientific methods would be available to properly assess such risks (cf. EU H2020 projects such as GRACIOUS, SUN, NanoReg, NanoReg2, ProSafe and many others). This has generated a myriad of assessment methods including, among others, (1) *in silico* modelling tools to predict physicochemical properties or hazard endpoints [5–8]; (2) control banding tools, which assign NMs to hazard and exposure bands [9], (3) grouping methodologies to enable read-across of data [5,6,10,11], (4) risk ranking tools, aimed at assigning a NM to a hazard or risk class (e.g. low, medium, high) [12,13]. The majority of these approaches are intended for use by industries for risk screening purposes and most of them do not address the requirements of the Regulation on Registration, Evaluation, Authorisation and Restriction of Chemicals (REACH).

The European Regulation on Classification, Labelling and Packaging of Substances and Mixtures (the CLP regulation) [14] implements the United Nations' Globally Harmonised System of Classification and Labelling of Chemicals (UN GHS [15]). The scope of the CLP regulation is to enable the identification of hazards posed by a substance or a mixture placed on the market, see section S1 of the Supplemental Information (SI), through the application of specific classification criteria to available hazard data, and then to provide the most appropriate hazard labelling and consequent information on safety measures. Classification, and subsequent steps, are normally undertaken by industry, and only for certain substances and mixtures shall the European Chemical Agency's (ECHA) committee on risk assessment propose a harmonised Classification and Labelling, in short C&L (REACH, Art 76). The classification is firstly based on the data submitted to REACH, which is generated according to test methods developed by the Organisation for Economic Co-operation and Development (OECD) and taken up by the EU Test Methods Regulation [16]. Hence for the application of the CLP regulation to NMs it is of utmost importance to ensure that available test methods are applicable to NMs. The OECD is currently confirming this applicability and updating its test methods [17].

REACH and the CLP regulation address 'substances' and apply to NMs as these are nanoforms of chemical substances as defined by REACH, which specifically outlines data requirements for nanoforms in its annexes [18]. NMs, however, are not specifically mentioned in CLP criteria, nor in the UN GHS [15,19]. IUCLID (International Uniform Chemical Information Database) is the software supporting the submission of registration dossiers according to the requirements of REACH and now includes fields to collect NM-related data [20]. In the European Union, ECHA hosts the inventory of classified substances. Some NMs, so-called next generation NMs [21], may pose new challenges under REACH [22] as for example it may not be straightforward to categorise them in the REACH categories ‘substance’, ‘mixture’ or ‘article’, impacting on the associated data requirements, and hence data availability for C&L. Nevertheless, possible issues related to classifying NMs according to the CLP regulation include that until updated or new OECD test methods for NMs are available, it may not be possible to generate the data for identifying hazards for NMs. This NM hazard data is needed as adverse effects of NMs may be different from the non-nanoform and between nanoforms depending e.g. on their physicochemical properties [23].

The UN has an informal working group[a] that is looking into the applicability of UN GHS criteria to NMs. In this context, the applicability to five human health endpoints using four NMs as examples has been evaluated. The conclusion was that in general the UN GHS is applicable to NMs with some limitations for certain NMs (e.g. NMs with relatively high specific surface area and low pour densities), for which it may not be technically feasible to test up to dose levels corresponding to the less severe hazard categories for acute toxicity and Specific Target Organ Toxicity (Repeat Exposure) [24]. Indeed, the World Health Organization recently provided a systematic review on classifying 11 NMs by applying UN GHS criteria [25].

To ensure a harmonised implementation of the criteria for performing C&L, ECHA’s CLP guidance [19] proposes several steps to arrive at a classification of a substance or a mixture: (1) identification of available information regarding the potential hazards of a substance or mixture; (2) examination of the information gathered to assess whether it is relevant, reliable and sufficient for classification purposes; (3) evaluation of the data by comparing the data to the classification criteria of the hazard classes; and (4) decision on whether the hazard information for the substance or mixture meets the criteria for one or more hazard classes, which if affirmative leads to the classification of the substance or mixture for those hazard classes.

Hazard information relevant for classification include experimental data generated in tests for physical hazards, toxicological and ecotoxicological tests, historical human data such as records of accidents or epidemiological studies, information generated in *in vitro* tests, (Quantitative) Structure Activity Relationships ((Q)SAR) and read across from other similar substances. In some cases, the classification decision may be straightforward, requiring only an evaluation of whether the substance gave a positive or negative result in a specific test that can be directly compared with the classification criteria. In other cases, for instance when the available data is heterogeneous or incomplete, a Weight of Evidence (WoE) approach combined with scientific judgement is recommended to make a classification decision [2].

---

[a] https://unece.org/DAM/trans/doc/2018/dgac10c4/UN-SCEGHS-36-INF35e.pdf

The WoE approach aims at integrating different sources of information to perform an assessment based on the strength of all available evidence, and is particularly beneficial when the information from single studies are not sufficient to fulfil an information requirement, or when individual studies provide different or conflicting conclusions [26–28]. Moreover, it allows weighting and integration of incomplete studies and/or studies that on their own are considered to be not reliable [26–28]. It is thus particularly indicated for handling (meta)data generated in the last decades that is publicly available in databases such as the eNanoMapper [29]. The (meta)data in these databases is often incomplete and heterogeneous [30], but still usable in the WoE approach, while most of the other approaches require and assume data to be complete and sufficiently reliable [31].

The WoE is applicable to classifying substances according to the CLP criteria, as an alternative to either selecting the most conservative (reliable) study, or to use statistics such as the geometric mean of the available (reliable) data [19]. The WoE has been indeed recommended in scientific literature as well as regulatory agencies for both risk assessment and classification of substances [27,28,32,33], including ECHA's CLP guidance [19]. It uses a combination of information from several independent sources to give sufficient evidence to fulfil an information requirement. The weight given to the available evidence may depend on factors such as the quality of the data, consistency of results, nature and severity of effects, and relevance of the information. The WoE approach requires use of scientific judgment and, therefore, it is essential to provide adequate and reliable documentation, thus allowing to present the results in a coherent and sound way. As a general principle, the more information is provided, the stronger the WoE is [27,34].

While the WoE is a widely used scientific concept, the way in which it is applied is very heterogeneous. Indeed, the term WoE can be found in the scientific literature with a variety of meanings, ranging from the purely colloquial use of the word to structured approaches to data integration and interpretation. Linkov et al. [35] proposed a taxonomy of WoE methods (which is reported in section S2 of the SI), ranging from mainly qualitative approaches to increasingly more quantitative ones.

In this manuscript, we propose a quantitative WoE methodology for classification of NMs according to the CLP regulation, which is based on the recommendations by ECHA [26] and the European Food Safety Authority (EFSA) [27]. WoE approaches have already been applied for the safety assessment of NMs, specifically for hazard ranking and screening [34,36,37], for hazard assessment [38–40], for ranking and prioritizing occupational exposure scenarios [41], and for the identification of NMs in consumer products [42]. However, quantitative WoE approaches have not yet been applied to directly support regulatory classification according to the CLP regulation and UN GHS.

The main objective of this paper is to fill this gap by proposing a 4-step methodology consistent with ECHA's CLP guidance [19] and to the EFSA Guidance on the use of the WoE approach in scientific assessments [27]. The four steps involve: (1) the collection of data related to the endpoint of interest (each "piece of data" is referred to as a Line of Evidence, LoE); (2) the weighting of such LoEs according to quality and completeness criteria selected in the context of GRACIOUS and coherent with the data quality criteria indicated in ECHA's guidance on information requirements and chemical safety assessment (IR&CSA) [28]; (3) the comparison and integration of LoEs, probabilistically assigning NMs to CLP classes; (4) the assignment, based on expert judgment, of the NM to one of the hazard classes. The last two steps of this methodology were implemented in an R tool, which is available both as a standalone package and as a web application. Taking into account the probabilistic nature of the approach, the tool allows to highlight the uncertainty in the analysis, thus providing an additional tool to assessors to reach a conclusion on the classification. To demonstrate the methodology and the tool, we tested them as a proof-of-concept against a CLP relevant endpoint, i.e., Aquatic Toxicity.

The methodology is described in section 2, while the online R tool and results of applying the WoE methodology with Aquatic Toxicity data are reported in section 3 of this manuscript.

## 2 Methods

The proposed WoE methodology involves the following four steps, which should be applied to each endpoint of interest as specified in ECHA's CLP guidance [19]:

- *Collection of data for one or more NMs related to a CLP-relevant endpoint of interest (e.g. Aquatic Toxicity, Acute Toxicity, Carcinogenicity, etc. The full list is reported in section S1 of the SI)* : Each study is referred as a Line of Evidence (LoE). This step is further described in section 2.1 of this manuscript.
- *Each LoE is weighted by a score calculated based on data quality criteria*: Selected criteria (such as relevance, reliability and adequacy) and corresponding weights are described in section 2.2.
- *Comparison and integration of the weighed LOEs for each NM*: A Monte Carlo approach is adopted to provide a quantitative integration of the weighted evidence, probabilistically assigning the NM to one of the hazard classes defined in the CLP regulation [14] and in the UN GHS [15]. This step is described in detail in section 2.3. Users applying this methodology should be aware that one of the possible outcomes of the approach is "No classification" for a NM (cf. section S3 of the SI), meaning that the identified hazards do not raise significant concern, or no hazards were identified either based on available data or as no data was available, and thus the NM does not need to be classified according to the CLP criteria and the available data.
- *Decision on the WoE*: After obtaining probabilistic results for each hazard class, experts should draw conclusions on the WoE based on the results obtained in the previous step, assigning each NM to a hazard class, as described in section 2.4.

To facilitate the analysis, a tool written in the R language was developed. This tool (described in section 3.1) enables the input of the list of weighted LoE, performs the classification following the decision trees described in the CLP regulation (decision criteria and trees are reported in section 3 of the SI), and produces as an output a table summarizing the probability for each NM to be assigned to CLP classes. Furthermore, taking into account the probabilistic nature of the approach, this tool allows the evaluation of uncertainty in results. Finally, it allows to automatically classify the NMs into CLP categories (including "no classification" for NMs identified as non-hazardous) by using two possible strategies, namely by either assigning each NM to the class with highest probability, or by assigning the NM to the class with nonzero probability that raises the highest concern.

## 2.1 Collection of relevant Line of Evidence

The first step of the approach consists of collecting data relevant to the investigated CLP endpoint from databases, the literature, or other sources. This results in a list of LoE for a NM, as each LoE corresponds to a separate study. The value associated to each LoE could be either quantitative (i.e. numeric values, possibly drawn from statistical distributions) or qualitative (e.g. "Known carcinogenic potential", "Presumed carcinogenic potential", "No carcinogenic potential") depending on the decision trees specified in ECHA's CLP guidance [19]. Examples of such decision trees on (eco)toxicological endpoints relevant to the CLP regulation are reported in the SI (section S3). In selecting the data, it is indeed important to ensure that they are comparable and match the requirements of the decision trees. Specifically, in case of quantitative data if the units of measure are different from what is indicated in the decision trees for each endpoint [19], it is necessary to convert data using proper conversion factors (e.g. converting from ng/L to mg/L, in the case of Aquatic Toxicity) .

Moreover, the test endpoint type and result should match with the endpoint type and result specified in the guidance [19]. For instance, in the case of Aquatic Toxicity, to evaluate chronic effects it is recommended to choose "NOECs[b] or corresponding ECx" [19]. In principle, depending on the test species and on the protocol, a different concentration could be used as ECx (e.g. EC10, EC20, EC25), however Beasly et al. [43] have recently concluded that the EC10 is a more appropriate surrogate for the NOEC than the EC20. Similar conclusions were drawn also by the International Centre for Pesticides and Health Risk Prevention (ICPS) and by EFSA [44].

## 2.2 Weighting the LoE based on Data Quality and Completeness criteria

After collecting relevant data, the second step of the WoE approach requires the assignment of weights to each LoE according to data quality and completeness criteria. Several approaches have been proposed to provide a quality [34,45–51] and completeness [30] assessment of hazard data, many of which are based on the methodology proposed by Klimisch et. al 1997 [45]. Starting from this knowledge, we decided to adopt the following state-of-the-art criteria to assess data completeness and quality, which are coherent with the data quality criteria indicated in ECHA's IR&CSA guidance [28]:

- *Data completeness*: which refers to the degree to which all required (meta)data in a data set is available.
- *Data reliability*: which measures if a study was conducted in a reliable manner.
- *Data relevance*: which measures if a study was conducted using standard protocols/procedures that are indeed relevant to identify the hazards related to the endpoint.
- *Data adequacy*: defining the usefulness of the data for classification purposes.

[b] NOEC: No observed effect concentration, ECx: the effect concentration at which x% effect (mortality, inhibition of growth, reproduction, etc.) is observed compared to the control group.

In the following subsections it is thoroughly described how such approaches are implemented in the methodology.

### 2.2.1 Assessment of data completeness

Our approach evaluates data completeness for an (eco)toxicological endpoint of interest, with respect to completeness of the provided information related to the physicochemical characterisation of the material and to the testing procedure and test conditions. This is based on the work of Comandella et al. [30] who proposed that for NM-specific physicochemical datasets, data completeness is evaluated for 11 measured physicochemical properties: i.e., crystallinity, composition, particle size, surface chemistry, particle shape, specific surface area, surface charge, surface hydrophobicity, dustiness, water solubility and density. This data are either required by REACH for nanoform identification or are recommended by ECHA as a basis for grouping [30]. In addition, for (eco)toxicological datasets the evaluation of data completeness covers the information related to the testing procedure (e.g. reference to the Standard Operating Procedure, the tested endpoint, the assay name, etc.) and test conditions (e.g. the adopted dispersion protocol and medium, the concentration, details on the cell lines and culture conditions, etc.).

Specifically, a Completeness Score (CS) is computed for each study (LoE) based on a checklist of properties and conditions acquired from eNanoMapper templates, which are a refined version of the NANoREG templates [52], with additional templates recently proposed by the GRACIOUS project [53]. These templates structure the information on assays in a set of MS Excel templates, developed by the Joint Research Centre of the European Commission and released under the Creative Commons Share-Alike license [52] (cf. Figure 1 for an example of such a template).

| Sample Information | | | | | | | | | | | | | | Nanoreg2 | | Size distribution (as measured with other techniques) | | | | | | | |
|---|---|---|---|---|---|---|---|---|---|---|---|---|---|---|---|---|---|---|---|---|---|---|---|
| Replicate number | NM ID code | NM chemistry (core) | CAS Number | Vial number | NM supplier | Material State | Use of dispersant | Dispersant reference | Sample Name | reporting organisation | operator | date of preparation | date of analysis | Endpoint (OECD list, and others) | Assay name | Dispersion protocol | Size Distribution | Size distribution analyse method | Dispersion medium | Mass concentration mg/mL | Diameter nm | in weight (w) or numer (n) | PDI |
| | NM-300K | NPO_1892 | | 7730 | NANoREG | | | 7730 | NM-300K | SINTEF | | | | Toxicity tc | OECD TG 201 | NANoREG | | | | | | | |

| Experimental Parameters - Start of the ecotoxicity test | | | | | | | | | | | | | | Experimental Parameters - End of the ecotoxicity test | | | | | | | | | | | RESULTS | |
|---|---|---|---|---|---|---|---|---|---|---|---|---|---|---|---|---|---|---|---|---|---|---|---|---|---|---|
| Nominal initial concentration mg/L | Effective initial concentration mg/L | Exposure medium analyzed | Hydrodynamic diameter in exposure media nm | PDI in exposure media | Z-pot in exposure media mV | Microscopy verification (PSD) | Quantification in exp media UV/Vis Wavelength nm | UV/Vis Total concentration in exposure media mg/L | ICP-MS Total concentration in exposure media mg/L | pH | T °C | $O_2$ concentration mg/L | Exposure time h | Hydrodynamic diameter in exposure media nm | PDI in exposure media | Z-pot in exposure media mV | Microscopy verification (PSD) | Quantification in exp media UV/Vis Wavelength nm | UV/Vis Total concentration in exposure media mg/L | ICP-MS Total concentration in exposure media mg/L | pH | T °C | $O_2$ concentration mg/L | Assay Protocol REF | EC25 mg/L | EC50 mg/L |
| | | Milli-Q Water | | | | | | | | 6 | 25 | 0 | 72 | | | | | | | | 6 | 25 | 0 | | | 0.01 mg/l |

**Figure 1: Example of data reporting template for an ecotoxicological study, partially filled with information related to a study.**

In detail, for each LoE the CS is computed for each relevant template related to the physicochemical characterization as the number of items (parameters) reported, divided by the number of items (parameters) required by the template. Mathematically, given the set of required templates for physicochemical endpoints, the Completeness Score of the i-th template in the set ($CS_{template_i}$) is computed as follows:

$$CS_{template_i} = \frac{\text{number of items available}}{\text{number of items required by template}} \qquad \text{(Equation1)}$$

These CSs of the 11 templates associated to the corresponding physicochemical parameters are then averaged, obtaining a score for the LoE related to the physicochemical characterization of the NM, i.e., $CS_{physchem}$:

$$CS_{physchem} = \frac{\sum_{i=1}^{i=11} CS_{template_i}}{11} \qquad \text{(Equation2)}$$

The Completeness Score associated to the template of the (eco)toxicological endpoint, $CS_{ecotox}$, is computed similarly to the completeness score of physicochemical templates (Equation1, Equation2), Finally, $CS_{physchem}$ and $CS_{ecotox}$ are averaged, thus obtaining an overall CS for a particular study (LoE) quantifying the completeness of the information related both to the physicochemical characterization of the NM and the characterization of (eco)toxicological study associated to the endpoint of interest:

$$CS = \frac{CS_{physchem} + CS_{ecotox}}{2} \qquad \text{(Equation3)}$$

### 2.2.2 Assessment of data reliability

Following the works of Card & Magnusson 2010 [46] and Fernandez-Cruz et al. 2018 [50], data reliability assessment is assessed by means of the ToxRTool [54], which is an excel spreadsheet aimed at determining the reliability of (eco)toxicological data, extending the approach originally proposed by Klimisch et al. 1997 [45] and adapting it to include nano-specific considerations and to highlight the robustness of particle characterization. As for Klimisch scores, the tool assigns data to the following categories [54] based on the information provided by the user:

- *Category 1 - Reliable without restriction:* Studies or data from the literature or reports which were carried out or generated according to generally valid and/or internationally accepted testing guidelines or in which the test parameters documented are based on a specific (national) testing guideline or in which all parameters described are closely related/comparable to a guideline method.
- *Category 2 - Reliable with restrictions*: Studies or data from the literature, in which the test parameters documented do not totally comply with the specific testing guideline but are sufficient to accept the data or in which investigations are described which cannot be subsumed under a testing guideline, but which are nevertheless well documented and scientifically acceptable.
- *Category 3 - Not reliable*: Studies or data from the literature/reports in which there were interferences between the measuring system and the test substance or in which organisms/test systems were used which are not relevant in relation to the exposure (e.g., unphysiological pathways of application) or which were carried out or generated according to a method which is not acceptable, the documentation of which is not sufficient for assessment and which is not convincing for an expert judgment. In a WoE approach, this data may be considered, as also specified in the ToxRTool.
- It does not assign to *Category 4* ("*Not assignable*"), as this assignment should be made by directly by the user.

### 2.2.3 Assessment of data relevance

Relevance [45,48,49] covers the extent to which data and the test are appropriate for hazard characterization. Specifically, this step ensures that the study is conducted using protocols and procedures that are relevant to identify the hazards related to the endpoint.

Specifically, the protocol used to generate data for the endpoint should be among the ones recommended in ECHA's CLP guidance [19], or it should be analogous to these protocols. In some cases, it is also possible to use data generated from other protocols, if those are considered by expert judgement being relevant to the task.

This results in three categories for relevance:

- *Category 1 – Relevant data*: data derives from protocols which are relevant to the endpoint according to the CLP regulation and ECHA's CLP guidance [19], or analogous to these protocols.
- *Category2 – Partially relevant data*: data derives from other protocols, which are considered relevant to the task by expert judgment but not recommended by ECHA's CLP guidance [19].
- *Category 3 – Not relevant data*: data is not relevant to the task, or the adopted protocol is not reported.

Data in Category 3 – Not relevant data, being not relevant to the task, is discarded. It is worth noting that results in Category 2 – Partially relevant data should be comparable to the data required by the decision trees defined by the CLP regulation and in the guidance. In other words, it should have the same unit of measure (in case of quantitative data) and be related to the same test endpoint (e.g. it is not possible to use EC20 values where EC50 values are required).

### 2.2.4 Assessment of data adequacy

Adequacy defines the usefulness of data for classification purposes. In our approach, this is evaluated by considering the nature of the test generating the data, namely greater weights were associated to the most reliable test for risk assessment purposes (i.e., *in vivo* tests), while lower weights were associated to *in vitro* and *in silico* studies.

This is consistent with ECHA's IR&CSA guidance [28] as well as with other similar WoE approaches, for instance the one proposed by Hristozov et al. 2014 [34], where authors additionally considered the exposure route. This is however not needed in our approach since the exposure route is already implicitly considered when selecting the endpoint.

### 2.2.5 Definition of weights

Based on the four criteria described in the previous sections, the proposed weights assigned to each LoE, defined on a scale [0, 1], are summarized in Table 1. These weights were adapted from previous studies [34,36], where available, but here we gave progressively lower weights in the higher categories (i.e., categories numbered 2, 3 and 4).

**Table 1: Definition of weights for the criteria illustrated in section 2.2.**

| **Criterion** | **Description** | **Weight computation** |
|---|---|---|
| **Data completeness** | Based on the Completeness Score (CS) introduced by Comandella et al. 2020 [30] for physicochemical data, extended to also include (eco)toxicological data. | Weight is computed for each LoE as the average between the CS of the physicochemical data, and the CS of the (eco)toxicological data (see Equation3). The resulting score ranges from 0 (no information provided) to 1 (all the information required by each template is provided). |
| **Reliability** | Based on RToxTool outcomes. It is worth noting that in WoE approaches it is accepted to include also data which is not | • Category 1 (Reliable without restrictions): 1 |
| | reliable (Category 3), with lower weight [19,28]. | • Category 2 (Reliable with restrictions): 0.5<br>• Category 3 (not reliable): 0.1<br>• Category 4 (not assignable): 0 |
| **Relevance** | The protocol used to generate data for the endpoint should be among the ones recommended in ECHA's CLP guidance [19], or it should be an analogue of these protocols. In some cases, it is also possible to use data generated by using different protocols, if those are considered by expert judgement to be relevant to the task. | • Category 1 (Relevant data): 1<br>• Category 2 (Partially relevant data): 0.3<br>• Category 3 (Non relevant data): 0 |
| **Adequacy** | *In vivo* studies are considered to be the most adequate, while *in vitro* and *in silico* studies are considered to be less adequate for regulatory purposes. | • *In vivo*: 1<br>• *In vitro* 0.3<br>• *In silico*: 0.1 |

Once the evaluation of weights has been performed for all LoEs, a final weight is computed as the average of the four weights. Mathematically, given the list of all the LoEs for a NM, let $LoE_i$ the i-th LoE in that list with associated value $v(LoE_i)$, and let $w_1(LoE_i)$ be the weight associated to the "Completeness" criteria (i.e., the CS calculated in Equation3), $w_2(LoE_i)$ be the weight associated to "Reliability" criteria (cf. Table 1), $w_3(LoE_i)$ be the weight associated to "Relevance" criteria (cf. Table 1), and $w_4(LoE_i)$ be the weight associated to "Adequacy" criteria (cf. Table 1), the final weight associated to the LoE, namely $w(LoE_i)$, is computed as:

$$w(LoE_i) = \frac{w_1(LoE_i) + w_2(LoE_i) + w_3(LoE_i) + w_4(LoE_i)}{4} \qquad \text{(Equation4)}$$

This weight is then used to integrate the LoE to obtain a final categorization, as described in the following section.

## 2.3 Integration of the weighted LoE per NM and classification in CLP categories

After computing weights for each LoE, weighted LoE are integrated by using the Monte Carlo approach. Namely, LoE related to a NM are sampled with replacement a specified number of times (e.g.: 100000 times, cf. section S4.1 of the SI) with probability equal to $w(LoE_i)$: the corresponding result value $v(LoE_i)$ is then extracted from the sampled LoE and classified according to the decision tree of the endpoint of interest. Then, for each sampled LoE, the decision tree related to the endpoint and the corresponding thresholds [19] are followed, thus probabilistically assigning the NM to a hazard class.

It is worth emphasising again that endpoint values associated to each LoE, namely $v(LoE_i)$, should be comparable with each other and more importantly should be comparable to the requirements of the decision trees specified in ECHA's CLP guidance [19]. This means that results with different unit of measure or data type (with the exception of what is explained in section 2.3.1), as well as results related to different endpoint types, should be discarded or converted to be comparable to the requirements of the decision trees defined by the guidance and the regulation.

### 2.3.1 Integration of LoE with different data types

Data may not always be available as numeric values for the endpoint of interest; it may happen for instance to have Confidence Intervals (CIs) of a distribution, or data may be qualitative. Hence, the following methodology is applied to retrieve the endpoint value for each simulation:

- If the endpoint value is deterministic, this value is assigned to the sampled LoE. Qualitative data is treated as deterministic values. Qualitative results may indeed be easily mapped to deterministic values to perform the analysis, and then mapped back to the original values.
- If the endpoint value is an interval, a value is sampled with uniform probability from the interval.
- If the endpoint value is provided as CIs or as mean and standard deviation of a statistical distribution, at each iteration a random value is sampled from the corresponding distribution.
- If a list of possible measured values is provided, a value from this list is sampled.

### 2.3.2 Output of the integration

The results of the integration of the weighted LoEs are presented as a table, where each row represents a NM, each column corresponds to a hazard class, and in each cell the probability for a NM to belong to a class is highlighted. This table, then, could be displayed in a user-friendly way as a heatmap or as a histogram. An example of a heatmap displaying results after applying the WoE methodology is presented in Figure 2.

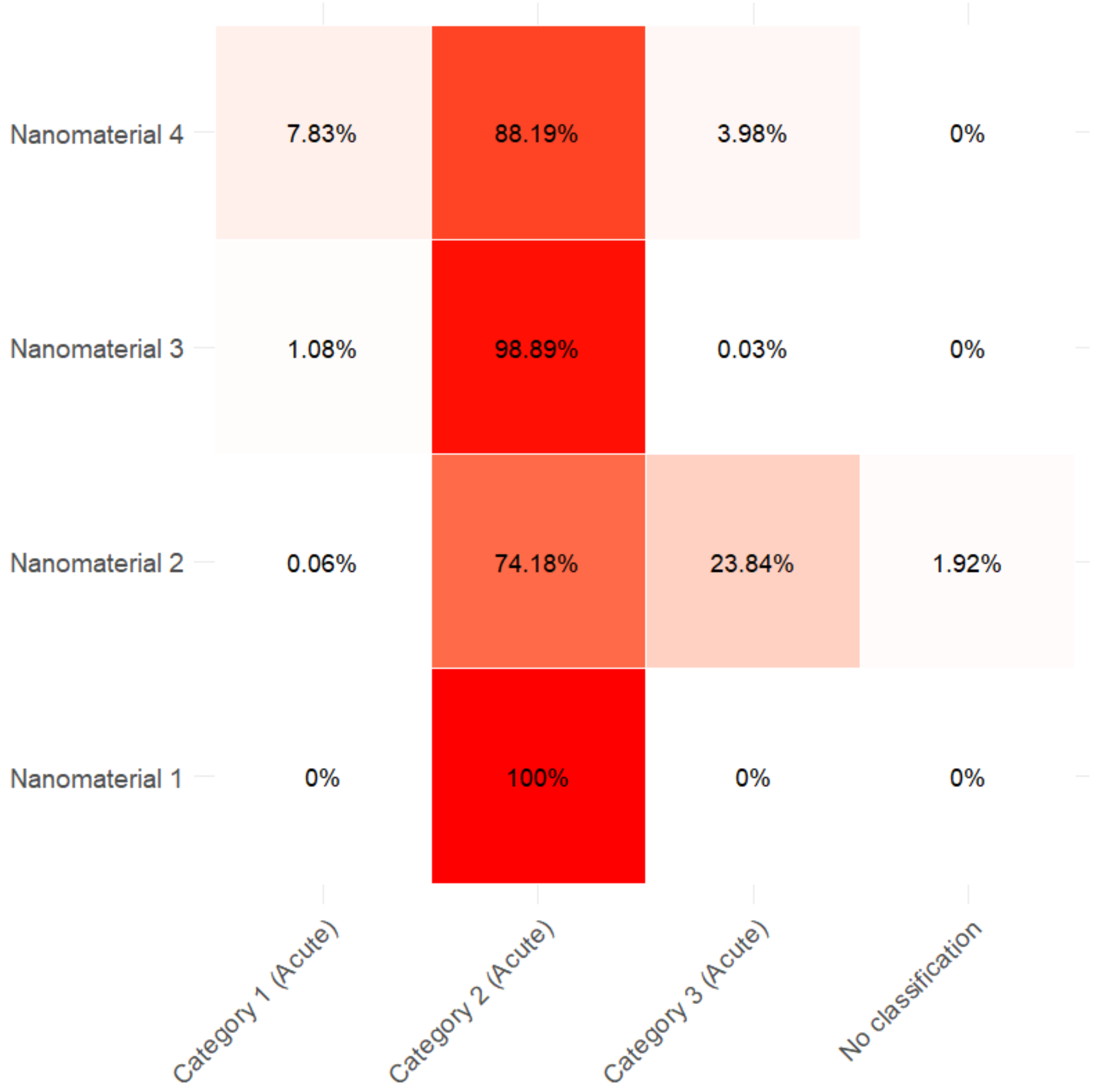

**Figure 2: Example of a heatmap displaying results of the methodology, generated with simulated data. The heatmap displays the probability for each NM of being assigned to a hazard class (according to CLP). Each box of the heatmap is coloured accordingly in a scale [white (0%) - red (100%)].**

Given the probabilistic nature of the methodology, this approach allows to easily evaluate the uncertainty of the results by simply repeating a sufficient number of times the approach (e.g., 10.000 times) and computing the 90% CI (i.e. the 5%-95% confidence interval) of the resulting probabilities of belonging to a class.

## 2.4 Assignment of NMs to a hazard class

The last step of the methodology involves a judgment on the results provided by the heatmap. The interpretation of this outcome allows experts to draw a conclusion, assigning for instance the NM to the most conservative category (i.e., the category that raises the major concern) with non-zero probability, or assigning it to the class with the highest probability. Moreover, when drawing conclusions on the WoE and thus in the final classification, experts should integrate this decision with considerations relative to the original data leading to the obtained results.

# 3 Results

The main result of this work is the WoE classification methodology described in section 2. Furthermore, we developed an online R tool implementing the last two steps of the methodology: i.e., integration of the weighted LoEs and decision on the classification, and we tested the methodology and the tool against a CLP relevant endpoint, i.e., Aquatic Toxicity. The tool is described in detail in section 3.1 and in the SI (section S4), while the results on the case study are presented in section 3.2.

## 3.1 R tool implementing the WoE methodology for classification

While evaluating a small set of weighted LoE for each NM may be straightforward, requiring only to consider the most conservative one or the one with the highest associated weight, when a large number of LoE are available, and especially when the associated data is not deterministic (cf. section 2.3.1), integrating this data and classifying the NMs of interest could become complicated. To facilitate the integration of the weighted LoE in order to support the classification decision (third and fourth steps of the WoE approach presented sections 2.3 and 2.4), we developed an R tool and made it publicly available at https://shinyapps.greendecision.eu/app/woe. This tool is capable of performing data integration and classification for more than one NMs simultaneously. Specifically, the tool requires as input an Excel file containing information on the weighted LoEs relevant to one or more NMs (more details on its structure are presented in the SI), the endpoint of interest, the number of simulations to perform to classify each NM, and the strategy to adopt to perform the automatic classification. Moreover, it allows to evaluate uncertainty, and in that case the number of simulations to evaluate uncertainty needs to be specified (a default value is available, i.e. 10000 simulations).

The tool produces as an output a table with the proposed classification and consequent hazard communication (according to the CLP regulation and the UN GHS), an image with the classification results (similar to the one in Figure 2), and if requested an image with highlighted the 90% Confidence Interval of the probability of each NM to be assigned to each hazard category.

A screenshot of the developed tool (Shiny application) is presented in Figure 3, while a user guidance, the technical implementation of the tool, and its performance evaluation are described in detail in section S4 of the SI.

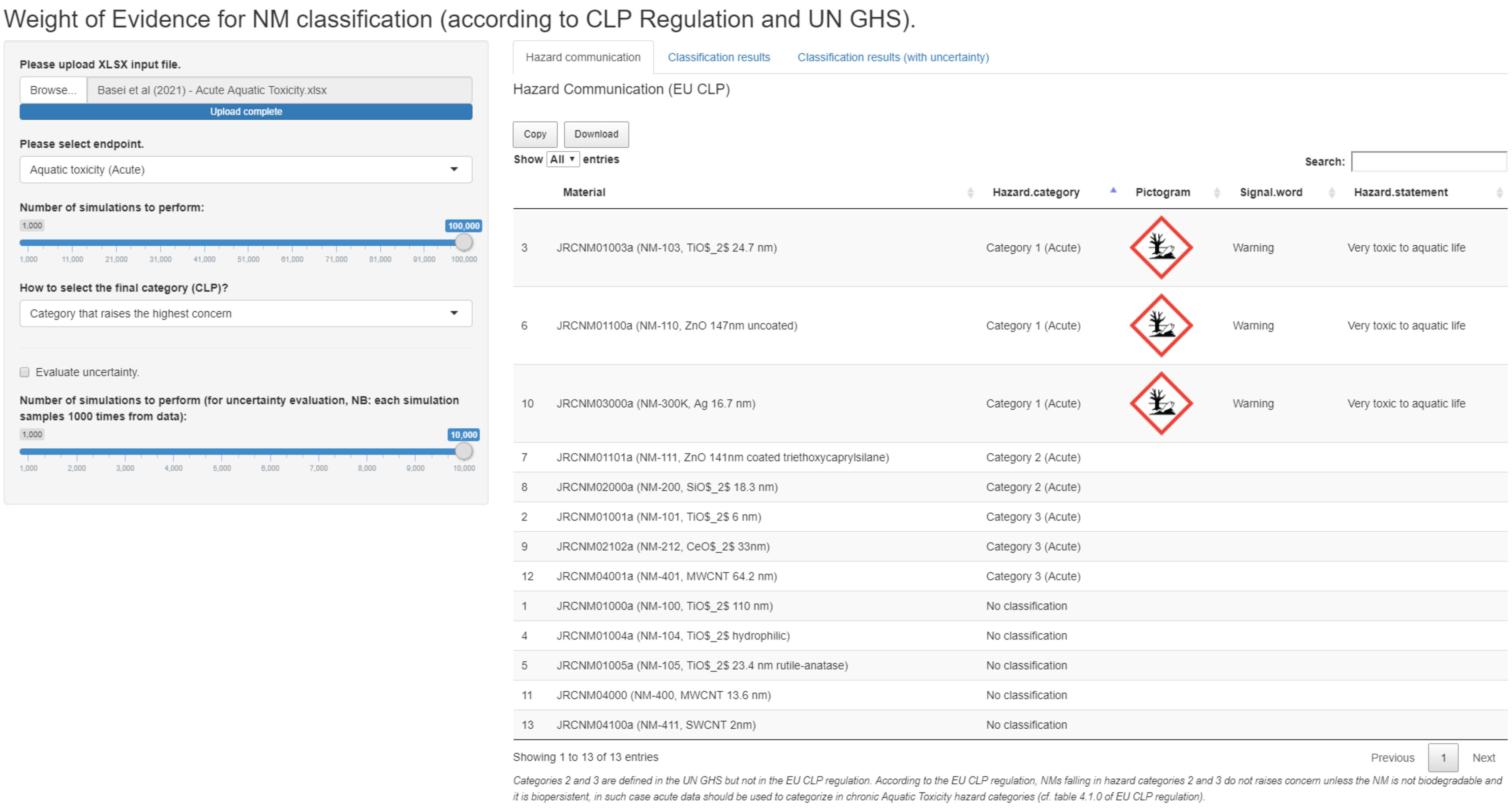

**Figure 3: Screenshot of the developed online R tool (Shiny application, publicly available at https://shinyapps.greendecision.eu/app/woe), highting the results of the proof-of-concept, described in section 3.2.**

### 3.2 Demonstration of the WoE approach in a case study: Aquatic Toxicity

To test the methodology, we extracted relevant data from the eNanoMapper database [29] selecting "Aquatic Toxicity" as the endpoint of interest. This endpoint was selected because it has recently been analysed in a classification approach identifying six groups according to different combinations of reactivity, morphology, and ion release [55]. The approach proposed by Hund-Rinke et al. [55], however, does not explicitly consider the classification criteria of the CLP regulation and is therefore not adapted to the regulatory requirements. Moreover, this endpoint was selected because a sufficiently large amount of aquatic toxicity data was available in the eNanoMapper database, and because the UN GHS [15] indicate three categories for acute Aquatic Toxicity and four categories for chronic Aquatic Toxicity. It is worth noting that the CLP regulation [14], and thus also ECHA's CLP guidance [19], do not include the need to classify the UN GHS Acute Aquatic Toxicity categories 2 and 3, cfr. section S3.10 of the SI. However, as these two categories are considered for Chronic Aquatic Toxicity if that is the available data and the substance is biopersistent and not biodegradable (cf. Table 4.1.0 of CLP regulation [14]), we decided to keep these two categories in the tool, and to highlight to users that the two categories do not in themselves trigger classification for acute hazards under the CLP regulation.

It is worth remarking, finally, that the information presented in this section should be considered as a proof-of-concept aimed at testing the methodology and the online R tool on real data, and not as a comprehensive classification study of the investigated NMs for regulatory purposes, as this is out of the scope of this manuscript.

In the following subsections the process of data collection and cleansing, and the application of the WoE methodology to the case study are described.

#### 3.2.1 Data collection and cleansing

The GRACIOUS instance of the eNanoMapper database includes data of the following European projects: GRACIOUS, ENPRA, MARINA, NanoGRAVUR, NANoREG, NanoTest, SANOWORK[c]. These data refer to three macro categories: physicochemical endpoints, toxicological endpoints and ecotoxicological endpoints. We extracted data from this database collecting a total of 81860 entries[d] (17810 of these related to physicochemical endpoints, 63655 related to toxicological endpoints, and 395 related to ecotoxicological endpoints) relative to 323 different NMs.

[c] Information on the mentioned projects may be found at the following web pages: https://www.h2020gracious.eu/ (GRACIOUS); http:// http://www.enpra.eu/ (ENPRA); https://cordis.europa.eu/project/id/263215 (MARINA); https://cordis.europa.eu/project/id/310584 (NANoREG); https://nanopartikel.info/en/research/projects/nanogravur/ (nanoGRAVUR); http:// http://www.nanotest-fp7.eu/ (NANOTEST); https://cordis.europa.eu/project/id/280716/ (SANOWORK).

[d] An entry is a single data element resulting from a query, expressing information on one specific quantity (or value) determined in an experiment using a specific measurement technique. For instance, a dose response curve of 5 elements resulting from a single study consists of 5 entries. Data was fetched on February 2020.

For testing the WoE approach, ecotoxicological data related to the aquatic compartment (313 entries) were identified. From this subset of data, we filtered out data not relevant to classify into Aquatic Toxicity categories (i.e., data not related to Aquatic Toxicity or data with test duration or test endpoints mismatching the requirements illustrated in the CLP regulation, cf. section S3.10 of the SI). Afterwards, we cleaned the resulting dataset as follows:

- Duplicates were removed and partial duplicates (entries referring to the same measurement but providing complementary information about it) were combined generating one individual result per study. Specifically, when both test endpoint value and its CIs were reported, we selected CI values, assuming lognormal distributions described by a 95% CI (i.e.: 2.5%-97.5% interval), consistently to what reported in the original source of most of the retrieved data [56].
- Erroneously reported information in the original data was corrected. Specifically, NM names were harmonized to be the same for the same NM in all entries, and information inserted in wrong columns was moved to the correct one.
- Empty columns, with the only exception of columns requested by the templates (cf. section 2.2.1), were removed.

The cleansing reduced the information to 47 entries of data generated by the NANoREG and MARINA projects, related to 13 NMs. These entries were related mostly to acute Aquatic Toxicity (45 entries), while only two studies addressed chronic Aquatic Toxicity.

### 3.2.2 Computation of weights

Once the data was collected and cleansed, we computed for each LoE weights for the four data quality criteria (i.e., completeness, relevance, reliability and adequacy) and integrated those into an overall weight for the LoE (as described in section 2.2). Then, two Excel spreadsheets were prepared to use as input to the online R tool, one for chronic Aquatic Toxicity and one for acute Aquatic Toxicity (such spreadsheets are attached to the manuscript).

### 3.2.3 Integration and classification of chronic Aquatic Toxicity data

Considering the available data for chronic Aquatic Toxicity, only one entry was available for each of two NMs, i.e. JRCNM01000a (NM-100, $TiO_2$ 110nm) and JRCNM01001a (NM-101, $TiO_2$ 6nm). In such cases the classification according to the CLP regulation is straightforward, as it is sufficient to follow the decision trees specified by ECHA's CLP guidance [19] to reach a classification conclusion. The tool was indeed developed to assist experts when more than one LoE are available for a NM. Nevertheless, it allows to perform the analysis also when a single LoE is available for a NM, thus for testing purposes we provided as input the spreadsheet, obtaining the results presented in Table 2. Such results indicate that for the two NMs no classification into chronic Aquatic Toxicity hazard is required, or, in other words, that according to the available data the two NMs do not raise concern for chronic Aquatic Toxicity.

**Table 2: Results of the application of the WoE methodology on chronic Aquatic Toxicity data. A dash indicates that the CLP regulation does not require to highlight in the label either a Pictogram, a Signal Word or a Hazard statement for the hazard category.**

| Material | Hazard category | Pictogram | Signal word | Hazard statement |
|---|---|---|---|---|
| JRCNM01000a (NM-100, $TiO_2$ 110 nm) | No classification | - | - | - |
| JRCNM01001a (NM-101, $TiO_2$ 6 nm) | No classification | - | - | - |

It is however important to emphasize that the tool (as specified in the SI, section 5) only classifies in categories 1, 2 and 3 with respect to chronic Aquatic Toxicity, as described in ECHA's CLP guidance [19]. When data does not allow classification in categories 1, 2 and 3, but there are some grounds for concern, Category 4 (Chronic) has to be considered. This category includes for example poorly water-soluble compounds for which effects are less than 50% at up to limit of solubility but which are not rapidly degraded, and which are biopersistent, unless other scientific evidence shows classification to be unnecessary [19].

### 3.2.4 Integration and classification of acute Aquatic Toxicity data

After evaluating the available data for chronic Aquatic Toxicity, we repeated the analysis for acute Aquatic Toxicity, by providing as input to the tool the spreadsheet generated as described in section 3.2.2. The results of the classification (selecting the class with nonzero probability that raised the highest concern) and the related hazard communication according to the CLP regulation, are presented in Table 3.

**Table 3: Results of the application of the WoE methodology with acute Aquatic Toxicity data. Categories 2 and 3 are defined in the UN GHS but not in the CLP regulation. According to the CLP regulation, NMs falling in hazard categories 2 and 3 do not raise concerns unless the NM is not biodegradable and it is biopersistent (cf. Table 4.1.0 of the CLP regulation [14]), in which case acute Aquatic Toxicity data should be used to categorize chronic Aquatic Toxicity hazard categories. A dash means that the CLP regulation does not require any highlight on the label, i.e. no Pictogram, Signal Word or Hazard statement for the hazard category.**

| Material | Hazard category | Pictogram | Signal word | Hazard statement |
|---|---|---|---|---|
| JRCNM01003a (NM-103, $TiO_2$ 24.7 nm) | Category 1 (Acute) | 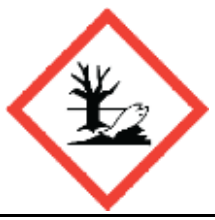 | Warning | Very toxic to aquatic life |
| JRCNM01100a (NM-110, ZnO 147 nm uncoated) | Category 1 (Acute) | 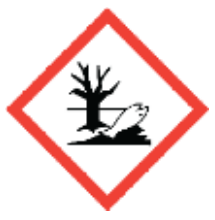 | Warning | Very toxic to aquatic life |
| JRCNM03000a (NM-300K, Ag 16.7 nm) | Category 1 (Acute) | 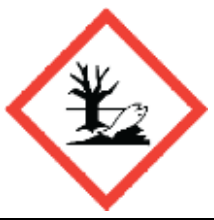 | Warning | Very toxic to aquatic life |
| JRCNM01101a (NM-111, ZnO 141nm coated triethoxycaprylsilane) | Category 2 (Acute) | - | - | - |
| JRCNM02000a (NM-200, $SiO_2$ 18.3 nm) | Category 2 (Acute) | - | - | - |
| JRCNM01001a (NM-101, $TiO_2$ 6 nm) | Category 3 (Acute) | - | - | - |
| JRCNM02102a (NM-212, $CeO_2$ 33nm) | Category 3 (Acute) | - | - | - |
| JRCNM04001a (NM-401, MWCNT 64.2 nm) | Category 3 (Acute) | - | - | - |
| JRCNM01000a (NM-100, $TiO_2$ 110 nm) | No classification | - | - | - |
| JRCNM01004a (NM-104, $TiO_2$ hydrophilic) | No classification | - | - | - |
| JRCNM01005a (NM-105, $TiO_2$ 23.4 nm rutile-anatase) | No classification | - | - | - |
| JRCNM04000 (NM-400, MWCNT 13.6 nm) | No classification | - | - | - |
| NM-411 (SWCNT 2nm) | No classification | - | - | - |

It is worth remarking that in the context of the CLP regulation, the UN GHS categories “Acute Aquatic Toxicity 2” and “Acute Aquatic Toxicity 3” are not mentioned (cf. section S3.10 of the SI and table 4.1.0 of the CLP regulation [14]). Nevertheless, if data related to bioaccumulation is available, and such data highlights that one or more of these NMs bioaccumulate in aquatic organisms, and/or data suggests that these NMs do not rapidly biodegrade, acute toxicity data should be considered for classification for chronic toxicity according to the CLP regulation (table 4.1.0 of the CLP regulation [14]).

### 3.2.5 Uncertainty analysis for acute Aquatic Toxicity

In the previous section, we decided to conservatively assign NMs to the class with nonzero probability that raised the highest concern. However, assessors may decide to choose the class with highest probability or may want to check the probability of each class before deciding on the assignment. In such cases the uncertainty assessment provided by the tool may be of help. This assessment is more meaningful when probabilistic data (i.e., probability distributions, ranges, etc.) is provided as value for one or more LoEs.

This was indeed the case for the analysis we made on acute Aquatic Toxicity (cf. section 3.2.4), where several studies were available for each material, and many of such had confidence intervals for the endpoint value. Thus, we also let the tool perform the uncertainty analysis, to investigate which were the ranges of the probability of assigning each NM to each hazard class. The results of uncertainty assessment are presented in Figure 4.

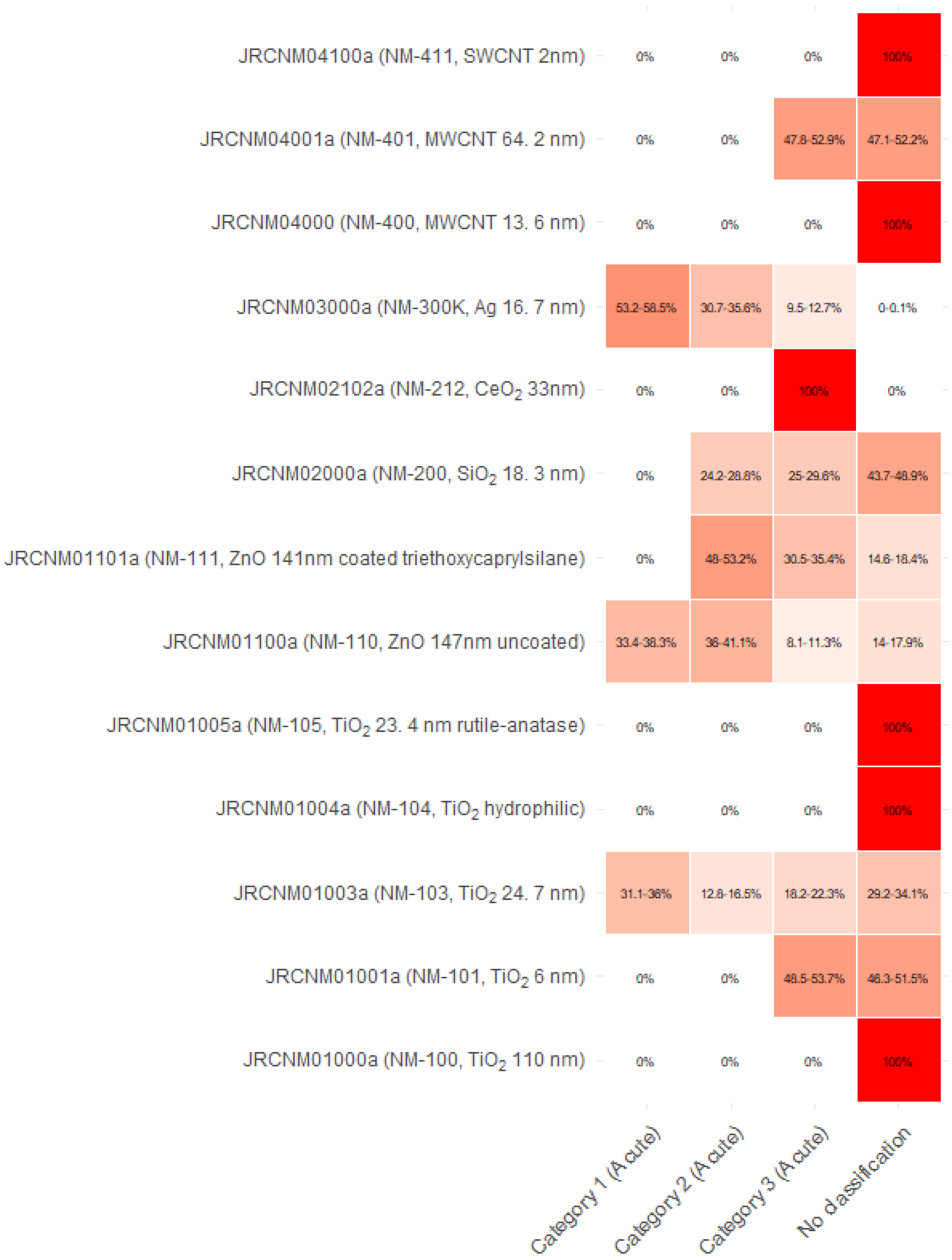

**Figure 4: results of the uncertainty assessment for acute Aquatic Toxicity data, displayed as a heatmap. In section 3.2.4, the NM was assigned to the category with nonzero probability that raised the highest concern, which is the most conservative category.**

By looking at uncertainty analysis results, assessors may assign JRCNM01100a (NM-110, ZnO 147nm uncoated) to Category 1, even though the material is associated to Category 2 with a higher probability, after observing that the confidence intervals of the probabilities are overlapping. Indeed, it is worth remarking again that expert judgment is always needed to provide a justification on the results of the classification in the last step of the methodology.

## 4 Discussion

In this manuscript, a quantitative WoE approach to probabilistically classify NMs according to the CLP criteria [14] was proposed. This approach involves four stages: (1) collection of data for different NMs related to the endpoint of interest (each study is referred to as a LoE); (2) computation of weighted scores for each LoE, based on state-of-the-art data quality and completeness criteria; (3) comparison and integration of the weighed LoEs for each NM through a Monte Carlo resampling; and (4) assignment of each NM to a hazard class. In addition, an online R tool was developed based on the WoE methodology, which allows to automatically perform the integration and classification of the weighted LoEs. Furthermore, to test the WoE approach, we selected Aquatic Toxicity as a case study endpoint, extracted and cleansed relevant data from the eNanoMapper database. Finally, the WoE methodology and the tool were tested against the cleansed data.

The proposed methodology accomplishes the main objective of this work, which is to propose and demonstrate a WoE approach for classification of NMs according to the requirements of the CLP regulation.

The main limitation of this approach was that data integration and comparison (step 3) could result being complicated when many LoE are available for the NM being analysed, and when the data are not deterministic. The developed online R tool is aimed at resolving this limitation by facilitating classification (step 4) according to two possible strategies: i.e., either selecting the most conservative category, or selecting the category with highest probability. Moreover, thanks to the probabilistic nature of the approach, the tool can provide risk assessors with information on the uncertainty in the classification, which can further aid them in the assignment of NMs to hazard categories.

An additional limitation of the proposed WoE approach is that in the computation of weights (step 2), expert judgment is needed to compute scores based on data quality and completeness criteria. Indeed, the obtained classification is strongly dependent on the choice of weights for each of the defined categories for data quality assessment, as demonstrated in the SI (section S6), where we vary the weights of each data quality category within the range [0,1]. There are ongoing efforts in the H2020 GRACIOUS and in the NMBP-13 projects (i.e., Gov4Nano, NANORIGO and RiskGONE) [e] to propose such criteria as part of a methodology that enables semi-automatic evaluation of data quality and completeness, including public workshops to define data quality criteria (such as the one organized by GRACIOUS project in June 2021[f]). Once this is achieved, it will be possible to seamlessly update the approach proposed herein to resolve this limitation.

[e] https://ec.europa.eu/info/funding-tenders/opportunities/portal/screen/opportunities/topic-details/nmbp-13-2018

To demonstrate the robustness of the proposed methodology, it was applied to an Aquatic Toxicity dataset extracted from the NANoREG project instance of eNanoMapper database. Specifically, chronic Aquatic Toxicity data related to two NMs were available, namely JRCNM01000a (NM-100, $TiO_2$ 110 nm) and JRCNM01001a (NM-101, $TiO_2$ 6 nm), and the application of our WoE methodology resulted in no classification for both NMs. In other words, neither JRCNM01000a nor JRCNM01001a raise concern for chronic Aquatic Toxicity. This is consistent with conclusions in the REACH registration dossier for nanoscale titanium dioxide (cf. https://echa.europa.eu/fr/registration-dossier/-/registered-dossier/15560/6/2/3).

For acute Aquatic Toxicity data related to 13 NMs (45 studies in total) were extracted from the eNanoMapper database. The application of the WoE methodology and the online R tool with these data generated the following results:

- Three NMs, namely JRCNM01003a (NM-103, $TiO_2$ 24.7 nm), JRCNM01100a (NM-110, ZnO 147nm uncoated) and JRCNM03000a (NM-300K, Ag 16.7 nm) were classified in Category 1, which means that they are of concern in terms of acute Aquatic Toxicity. Similar conclusions can be found in the literature, where Ag [57,58] and ZnO NMs [59] are known to be “extremely toxic” to aquatic organisms [60]. However, it is worth noting that while in the REACH registration dossier, nanosized $TiO_2$ are generally considered not to be acutely toxic to aquatic organisms (cf. https://echa.europa.eu/fr/registration-dossier/-/registered-dossier/15560/6/2/1), there are studies suggesting toxic effects of JRCNM01003a on algae in concentrations under 1 mg/L [61–63]. Indeed, the LoE we extracted from the eNanoMapper database that led to the classification of this material in Category 1 was related to a study on the algae *P. subcapitata* [56]. Possibly the aluminium coating of JRCNM01003a could be a confounding factor [64], and indeed in the original study a high content of impurities is reported which may have led to the low EC50 value (i.e., the concentration at which the algae growth inhibition is 50%), while adding the SR-NOM (Standard Suwannee River Natural Organic Matter) improved the stability of the dispersions and alleviated the adverse effects completely [65]. This result could also be explained by the fact that $TiO_2$ NMs may adsorb nutrients necessary for algae growth in closed artificial systems, while in natural environments this does not happen, as also pointed out both in the original study [65] and in the REACH registration dossier (https://echa.europa.eu/fr/registration-dossier/-/registered-dossier/15560/6/2/6). Hence, after evaluating this additional information, assessors may conclude that JRCNM01003a is not "extremely toxic for the aquatic environment".

[f] https://www.h2020gracious.eu/event/assessing_quality_and_completeness_of_nanosafety_data_for_risk_assessment_purposes

- JRCNM01101a (NM-111, ZnO 141 nm coated with triethoxycaprylsilane) and JRCNM02000a (NM-200, $SiO_2$ 18.3 nm) were classified in Category 2, which is defined in the UN GHS but not triggering classification under the CLP regulation. These results are consistent with the studies originally generating the data [56]; moreover there are studies in the literature highlighting dose-dependent ecotoxicological effects of $SiO_2$ NMs on crustacea *D. magna* [66] and on microalgae [67], and as aforementioned the ZnO NMs are reported to be generally toxic in the aquatic environment [59,60]. Furthermore, the decrease of toxic effects for the coated ZnO is confirmed by other studies [68], which explains why this modified material is in Category 2, while the uncoated ZnO is classified as more toxic Category 1 material.
- Three NMs, namely JRCNM01001a (NM-101, $TiO_2$ 6 nm), JRCNM02102a (NM-212, $CeO_2$ 33nm) and JRCNM04001a (NM-401, MWCNT 64.2 nm) were classified in Category 3, which is defined in the UN GHS but not triggering classification under the CLP regulation. These results are consistent with the published literature. Similar conclusions to what is already discussed for other $TiO_2$ nanoforms could be made for JRCNM01001a, while for the $CeO_2$ NMs no significant acute Aquatic Toxicity is reported in the literature and only chronic Aquatic Toxicity effects were observed [69]. Similarly, the available studies in the literature demonstrate that the MWCNTs are not concerning with respect to acute Aquatic Toxicity [70].
- Finally, three nanoforms of $TiO_2$ (namely JRCNM01000a, JRCNM01004a and JRCNM01005a) and JRCNM04000 (NM-400, MWCNT 13.6 nm) did not require categorization neither under CLP nor under UN GHS, or in other words are not of concern in terms of acute Aquatic Toxicity. Similarly, Carbon Nanotubes (NM-411, SWCNT 2 nm) were not categorized in any of the hazard classes related to acute Aquatic Toxicity, although studies categorizing these NMs in Category 3 could be found in the literature [71], which confirms that for a more reliable categorization more data may be needed.

The results obtained in our case study are consistent with the outcomes of recent studies from the published literature, which confirms the validity and the robustness of the proposed WoE methodology. However, it is important to stress that these results should only be considered as a proof-of-concept aimed at testing our approach and the online R tool with real data, not as classification results for regulatory purposes. To perform classification for regulatory purposes, it may be necessary to acquire stronger evidence based on more data of high quality.

## 5 Conclusions

In the GRACIOUS project, we proposed a four-step quantitative WoE methodology to classify NMs according to the requirements of the CLP regulation. The methodology allows the weighting of the available information according to data quality and completeness criteria results from heterogeneous studies, integrating such studies, and then evaluating the results to obtain a classification. Moreover, the probabilistic nature of the proposed approach enables highlighting the uncertainty in the analysis. To facilitate the implementation of this methodology, we developed an R tool, which was made publicly available both as a standalone package and a Shiny application (https://shinyapps.greendecision.eu/app/woe).

To demonstrate the robustness of the methodology and the tool, a proof-of-concept exercise against real data related to the CLP endpoint “Aquatic Toxicity” was performed. The obtained classification results were consistent with conclusions made in the projects (i.e., NANoREG and MARINA) and published studies that originally generated the data used in our study and to the published REACH registration dossiers. This confirmed the validity and the soundness of the obtained results and the overall robustness of our approach.

## Conflict of interest statement

We declare that we have no financial and personal relationships with other people or organizations that can inappropriately influence our work, there is no professional or other personal interest of any nature or kind in any product, service or company that could be construed as influencing the position presented in, or the review of, the manuscript entitled.

## Acknowledgements

The GRACIOUS project has received funding from the European Union’s Horizon 2020 research and innovation programme under grant agreement No 760840.